\documentclass[twocolumn,twoside,slac_two]{revtex4}
\usepackage{graphicx}
\usepackage{fancyhdr}

\def\apj{\emph{ApJ.}}
\def\apjl{\emph{ApJ. Lett.}}
\def\aap{\emph{A.\& A.}}
\def\araa{\emph{ARA\&A}}

\def\nat{\emph{Nature}}

\def\physrep{\emph{Physics Reports}}

\begin{document}

\title{A Proposal to Localize Fermi GBM GRBs Through Coordinated Scanning of the
GBM Error Circle via Optical Telescopes}

%

\author{T. N. Ukwatta, J. T. Linnemann, K. Tollefson, A. U. Abeysekara}
\affiliation{Department of Physics and Astronomy, Michigan State
University, East Lansing, MI 48824, USA.}
\author{P. N. Bhat}
\affiliation{University of Alabama in Huntsville Center for Space
Plasma and Aeronomic Research, 320 Sparkman Dr. Huntsville AL,
35805, USA.}
\author{E. Sonbas} \affiliation{University of Ad{\i}yaman, Department of Physics, 02040
Ad{\i}yaman, Turkey.}
\author{N. Gehrels} \affiliation{NASA Goddard Space Flight Center, Greenbelt,
MD 20771, USA.}

\begin{abstract}
We investigate the feasibility of implementing a system that will
coordinate ground-based optical telescopes to cover the Fermi GBM
Error Circle (EC). The aim of the system is to localize GBM
detected GRBs and facilitate multi-wavelength follow-up from space
and ground. This system will optimize the observing locations in
the GBM EC based on individual telescope location, Field of View
(FoV) and sensitivity. The proposed system will coordinate GBM EC
scanning by professional as well as amateur astronomers around the
world. The results of a Monte Carlo simulation to investigate the
feasibility of the project are presented.
\end{abstract}

\maketitle

\thispagestyle{fancy}


\section{Introduction}
Gamma-ray bursts (GRB) are bursts of gamma-rays that arguably
signal the birth of a black hole somewhere in the universe. Based
on the duration and spectrum, two classes of bursts have been
observed~\citep{Kouveliotou1993}: those that last less than two
seconds and have on the average hard spectra (short GRBs), and
those that last longer than two seconds and are spectrally softer
(long GRBs). The exact nature of the GRB progenitors is unknown,
although it is possible that long GRBs come from the collapse of
massive, rapidly rotating stars~\citep{Woosley2006,
WoosleyBloom2006} and short GRBs result from the merger of compact
objects~\citep{Eichler1989, Narayan1992}. Regardless of the
progenitor system, accretion onto the resulting compact object is
thought to create a highly relativistic jet. The prompt gamma-ray
emission from the GRBs may arises from the internal shocks due to
collisions of faster shells with slower ones ejected earlier by
the central engine. The subsequent softer multi-wavelength
emission, referred to as the afterglow, may be due to the
collision of the fireball with the extra-stellar
material~\citep{Rees1994,Piran1999}.

Our understanding of GRBs progressed very rapidly after the
detection of multi-wavelength afterglows. Well localized,
favorably positioned GRBs get fairly good multi-wavelength
afterglow coverage. Currently, the leading GRB afterglow detection
mission is $Swift$ which detects 90--100 GRBs annually. Most of
the $Swift$ GRBs get observed by various instruments around the
world because of its rapid arc-minute localization capability.
Compared to $Swift$, $Fermi$ Gamma-ray Burst Monitor (GBM) detects
about 250 burst per year but with poor localization. The Error
Circle (EC) of $Fermi$ GBM detected bursts is too large for a
single telescope to observe effectively. The typical statistical
uncertainty of the GBM burst location is about 3.3 degrees.
However, when combined with the systematic uncertainty of 3.8
degrees~\citep{Briggs2009}, the total burst location uncertainty
is $\sim$ 5.0 degrees (i.e., 5.0 degree error radius). Naturally,
a brighter burst will have a smaller GBM EC than a weaker burst.

Even though the localization is poor, GBM detected bursts have
very good timing and spectral information including crucial
$E_{\rm peak}$ measurements ($E_{\rm peak}$ is the peak energy of
the GRB $\nu F_{\nu}$ spectrum). If there is a method to localize
GBM detected GRBs to a few arc-seconds uncertainty, then large
telescopes can do deeper follow--up observations to determine the
redshift of the burst and also potentially identify any emerging
supernova. In addition, $Swift$ can also slew quickly to the GBM
burst in order to observe the X-ray afterglow and obtain its light
curve in X-ray wavelengths.

Based on $Swift$ observations about $\sim$ 60\% of GRBs have
optical counterparts~\citep{Gehrels2009}. These optical
counterparts are detected by various observatories with R
magnitudes ranging from 14 to 22 within few hours after the
burst~\citep{Fiore2007}. Thus, it is reasonable to assume about
60\% of the GBM detected GRBs also have optical counterparts with
similar brightness distribution. If we were able to cover the
entire GBM EC within about 24 hours after the burst it is
conceivable that we would be able to find optical afterglows of
$\sim$150 GRBs per year, which is more than the total number of
burst $Swift$ detects per year.

Due to the small energy range (15-150 keV) of the $Swift$ Burst
Alert Telescope (BAT), $Swift$ measurements alone cannot constrain
the $E_{\rm peak}$ of all BAT detected
bursts~\citep{Sakamoto2009}. In contrast, due to the wide energy
range (8 keV - 40 MeV) of GBM, all GRBs detected by GBM have
fairly good $E_{\rm peak}$ measurements. Hence, addition of
possibly another $\sim$100 bursts per year with good $E_{\rm
peak}$ and redshift measurements may allow us to explore the
validity of various GRB luminosity relations and to conduct
detailed GRB Hubble diagram studies.

We have investigated the feasibility of using a system to do
coordinated monitoring of the BAT field-of-view (FoV) for prompt
optical emission from GRBs~\citep{Ukwatta2011}. The study showed
that with the current instrumentation, performing such a
coordinated monitoring is not practical mainly due to the BAT's
very large FoV. However, a similar coordinated observing campaign
can be used to find the optical afterglow of GBM detected bursts.
The GBM EC is much smaller than the BAT FoV and observers do not
need to continuously monitor the field to detect the optical
afterglow. This enables a given observatory to perform multiple
observations inside the GBM EC and thereby increase the chance of
a afterglow detection.

\section{Method and Feasibility}

The basic proposal is to design a system to facilitate scanning of
the GBM EC for optical emission from GRB afterglows. This
observing program will be specially aimed at amateur astronomers
around the world. Proliferation of amateur telescopes with high
quality CCD cameras has opened a new avenue to study optical
emission from GRBs. The basic objective of the system is to
coordinate a significant number of ground based telescopes to scan
different patches of the GBM EC in order to find the location of
the optical afterglow. Unlike the GCN system~\citep{Barthelmy1998}
which sends notices to large number of recipients, this system
will send customized targeted messages to individual registered
telescopes. These individual messages will be sent via email or
socket connections and they will have one or more assigned
pointing locations for each telescope. The target telescopes can
be either robotic or non-robotic. The selection of various patches
in the GBM EC will be done based on the number of available
telescopes, individual telescopes' physical location, Field of
View (FoV) and sensitivity. It is also reasonable to assume that
these telescopes can observe multiple patches of the GBM EC, which
will increase the chances of detection significantly. 

Some of the important impacts of the proposed project are:
\begin{enumerate}
    \item The project will significantly enhance the value of GBM as a
GRB discovery instrument.
    \item Potentially increase the number of
burst with good timing, spectral and redshift measurements.
    \item The project will allow and attract the participation of amateur
astronomers and their telescopes.
\end{enumerate}

\begin{figure}[t]
\centering
\includegraphics[width=0.5\textwidth]{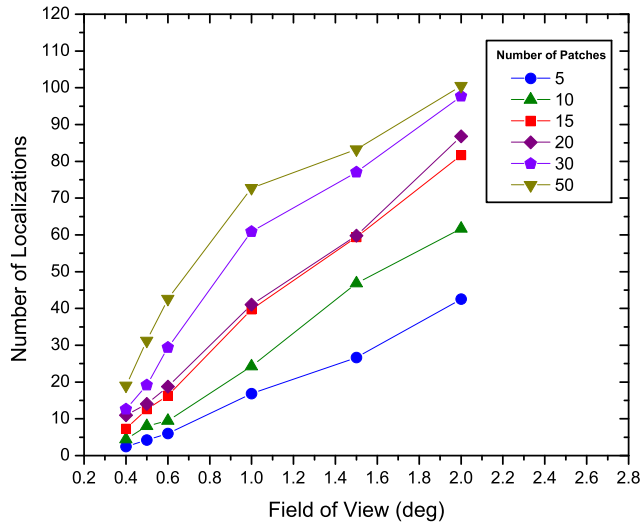}
\caption{The number of localized bursts as a function of the FoV
for 25 participating telescopes. Various curves corresponds to
different number of observations (or number of patches) a given
telescope can perform.} \label{sim_results01}
\end{figure}

\begin{figure}[t]
\centering
\includegraphics[width=0.5\textwidth]{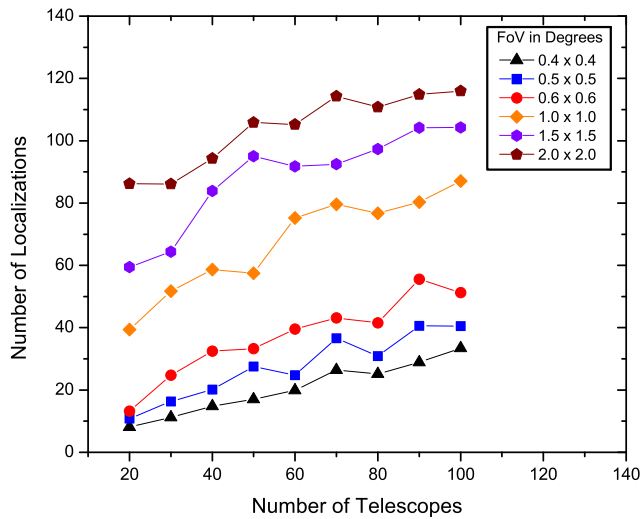}
\caption{The number of localized bursts as a function of number of
participating telescopes in the program assuming each telescope
can observe 20 patches.} \label{sim_results02}
\end{figure}

In order to investigate the feasibility of the project, we
performed a Monte Carlo simulation to study the probability of
detection of optical afterglows from $Fermi$ GBM GRBs. We assumed
that GBM detects about 150 GRBs with optical afterglows (total
rate is 250 per year) per year distributed isotropically in the
sky and throughout the year. We have distributed telescopes in
such a way that they roughly trace the major cities in the world.
Then for each burst we tracked the path of the Sun and selected a
set of telescopes away from the Sun and within few hours from the
burst location to scan the GBM EC. We also calculated the
illuminated fraction of the Moon's disk ($f_{\rm moon}$) at the
time of each burst. The probability, $P$, of finding a GRB was
estimated by
\begin{equation}
P = 1 - (1-p)^n.
\end{equation}
Here $n$ is the total number of independent attempts to observe,
with probability of success, $p$. In this case $n$ is equal to the
product of number of available telescopes and number of patches
each telescope can observe. We calculated the probability of
success, i.e., the probability of detecting a given burst
afterglow per observation using the following equation.
\begin{equation}
p = \frac{\textrm{Tel. FoV Solid Angle}}{\textrm{GBM EC Solid
Angle}} \times (1.0 - f_{\rm moon}) \times 68 \textrm{\%}.
\end{equation}
Note that typically the one $\sigma$ GBM EC is $10.0^{0} \times
10.0^{0}$. We repeated this procedure for every simulated burst,
while changing the total number of telescopes participating in the
program, the FoV of telescopes, and total number of patches a
given telescope can cover.

The results of our simulation are shown in
Figure~\ref{sim_results01} and Figure~\ref{sim_results02}. For
these particular simulations we have assumed that all the
telescopes have the same FoV and all telescopes can cover some
constant number of patches in the GBM EC. Furthermore, we assumed
that these telescopes will be able to observe assigned sky patches
within a few hours after the burst. Hence, in the simulation we
used only telescopes which are within a few hours ($\sim$ 6 hours)
of the burst location. The number of participating telescopes in
the observing program was varied from 20 to 100.

\begin{figure*}[htp]
\begin{center}
\centering
\includegraphics[width=0.9\textwidth]{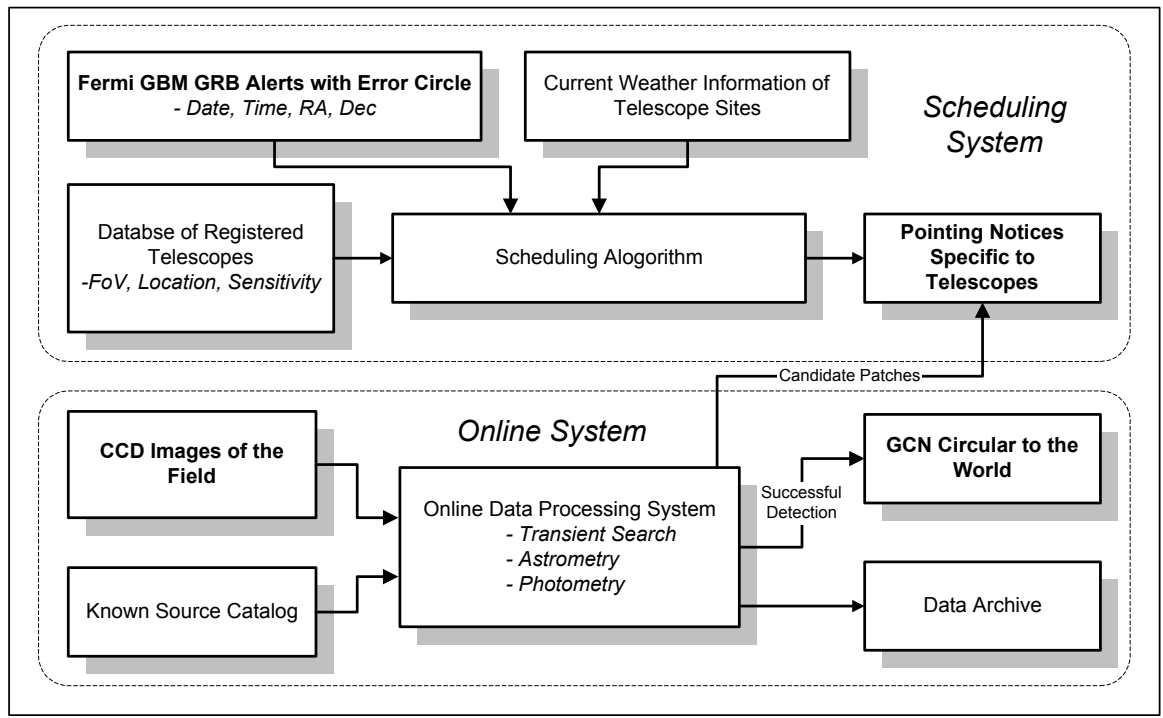}
\caption{Schematic block diagram of a potential system to localize
GRB bursts.}\label{diagram}
\end{center}
\end{figure*}

Figure~\ref{sim_results01} shows the number of localized bursts
($P\, \times$ Total Number of Bursts) as a function of the FoV of
participating telescopes. Here we have fixed the number of
participating telescopes to 25. Various curves correspond to
different number of patches that each telescopes can observe.
Figure~\ref{sim_results02} shows the number of localized bursts as
a function of number of participating telescopes in the program
assuming that each telescope can observe 20 patches. The six
curves shown in the plot correspond to various field of views.
According to the simulation, with 60 telescopes participating it
is possible to detect about 40 GRB optical afterglows per year
using telescopes with FoV of $0.6^{0} \times 0.6^{0}$. This value
is close to the value of a typical FoV of an amateur telescope. It
is also interesting to note that if we have about 10 telescopes
participating with FoV of $1.0^{0} \times 1.0^{0}$, then it is
possible to detect about 25 GRB afterglows per year. In order to
put these values into perspective we point out that thus far, no
one has managed to observe optical afterglow of a GRB based only
on a GBM localization (the GBM has been detecting GRBs for more
than three years).

It is also worth noting that not all the GBM locations will have a
statistical error of 3.3 degrees. About a third of the GRBs will
have a statistical error less than this value. About 10\% will
have error of 1 degree or less. In such cases the GBM EC radius
will be less than 5 degrees and may be as small as 4 degrees.
Obviously, for those cases we have a much higher chance of
detecting the afterglow.

A schematic block diagram of a potential software system is shown
in the Figure~\ref{diagram}. The system has two components: 1) a
Scheduling System that will assign various observing patches to
participating telescopes, and 2) a Online System that will let
observers to upload their images and search for candidate
transients.

The algorithms in the Scheduling System will check the GBM EC
observability of each participating telescope and assign them to
different parts of the GBM EC. In doing this the algorithm will
consider individual telescopes' FoV, sensitivity and local weather
conditions. In addition, it will also assign more than one patch
for each astronomer. On average an amateur astronomer may receive
about 40 notices per year. The exposure time for each patch
depends on many factors such as aperture, seeing, type of CCD
camera etc and typically may vary from 1 to 30 mins. The
probability of success depends on the telescope configuration
(FoV, sensitivity), local weather and sky conditions, and the
number of patches observed. However, every amateur observer who
submits an observation to the system will get credit for their
effort by being a co-author of the subsequent GCN notice that
results from a successful detection.

The Online System is envisioned to have a web interface where the
participants can submit their observations. It will also have
online tools that will compare the submitted observations with
existing catalogs and search for the optical afterglow of the GRB.
If one of the observations has a positive detection then the
system will initiate a follow-- up observation to establish
whether the candidate source is fading. If the candidate is found
to be fading (telltale signature of a GRB afterglow) then the
magnitudes of the two images will be determined and a GCN circular
will be sent.


\section{Summary and Conclusions}
We investigate the feasibility of implementing a system that will
coordinate ground based telescopes (both amateur and professional)
to scan the GBM EC in order to localize GBM bursts. Unlike the GCN
system, proposed system will send individual customized messages
to telescopes to observe certain patches in the GBM EC. The
scientific objective of the system is by localizing GBM detected
burst, we will be able to increase the number of GBM bursts with
mutil-wavelength followups potentially with redshifts
measurements. These measurements are scientifically very important
because there are hints that GBM bursts may represent
significantly different burst population. Based on our simulation,
we can detect about 25 GRB afterglows per year using just 10
telescopes with $1.0^{0} \times 1.0^{0}$ field-of-view. With more
telescopes participating in the program, we should be able to
detect many more afterglows and study a potentially interesting
burst population that is currently inaccessible to the GRB
community.

%

\end{document}